\documentclass[aps,twocolumn,superscriptaddress]{revtex4} 
\usepackage{graphicx}  
\usepackage{dcolumn}   
\usepackage{bm}       
\usepackage{amssymb}  
\usepackage{subfigure}
\usepackage{epstopdf}
\usepackage{comment}
\usepackage{color}
\usepackage{slashed}
\usepackage[utf8]{inputenc}
\usepackage{multirow}
\usepackage{footnote}
\usepackage{amsmath}
\allowdisplaybreaks[4]
\usepackage{accents}
\newlength{\dhatheight}

\usepackage{hyperref}

\def\figureautorefname~#1\null{Fig.\,#1\null}
\def\tableautorefname~#1\null{Tab.\,#1\null}

\def\equationautorefname~#1\null{Eq.\,(#1)\null}

\begin{document}
\title{ Partial wave analysis of two-body decay with helicity formalism}
\author{ Hang Chen  }
\email{xing-che17@mails.tsinghua.edu.cn }
\affiliation{Department of Physics, Tsinghua University, Beijing 100084, China}

\begin{abstract}
In this paper, We find that the angular distribution of two-body decay and scattering process that can be studied by introducing the helicity method.It has been argued that the angular distribution only depends on  magnitude of the decaying amplitude which final particle momentum is along the $Z$ axis. Therefore, when calculating the angular distribution,we only need to know the helicity amplitude in the $Z$ direction. At the same time, we also discuss the symmetry of physical processes, such as parity and time inversion transformation and so on. The effects of these symmetric transformations on the helicity state and the canonical states\cite{1993Helicity} are studied in this paper. When it is applied to two body decay process, the symmetry of the amplitude can be obtained by helicity method. In addition, we also introduce the density matrix to study the polarization of the final state particles, and discuss some inherent symmetries of the density matrix\cite{Chung1998General,1999Wavefunctions,2007Helicity,2008Covariant,1964Relativistic}. When we study the polarization and angular distribution of final state particles, we can use these properties to simplify density matrix without knowing all the elements of the density matrix.Then we  find that there is a close relationship between the partial wave\cite{2003Tree}coupling constants and the angular distribution parameters, indicating that these coupling constants can be obtained by experimental fitting to data.In a word, the properties of two body decay and two body cascade decay process can be analyzed by helicity method\cite{2014Unitarity}, which has a very good application .
\end{abstract}
\maketitle

\section{Partial wave analysis}
\indent We analyze the relationship of one or two-particle helicity and canonical states\cite{1965Systematics,2017Coherent}.These two kinds of physical states can be defined in canonical and helicity coordinate.These two kinds of coordinates can be related by a rotation with polar angle just the same as the final state momentum.\\
\indent So the state $|\Omega m_{1}m_{2}>$ equal to $|\Omega \lambda_{1},-\lambda_{2}>$.So the helicity state of two particles can be represented as follows,in which $a$ is a normalization constant to be determnated later.$|00\lambda_{1}\lambda_{2}>$ is a two particle helicity state whose momentum is along $Z$ axis.
\begin{equation}
|\Omega \lambda_{1}\lambda_{2}>=aU[R(\phi,\theta,0)|00\lambda_{1}\lambda_{2}>
\end{equation}
For two particles with definite total angular momentum $J$ and $M$.The partial wave analysis \cite{2003Helicity}require the helicity amplitude can be decomposed into the following form.
Consider that the initial state decay of $|jm>$ is  respectively$\lambda_ {1} $and $\lambda_ {2}$ In the case of the two particle helicity state of , the helicity amplitude is:
\begin{equation}
A_{\lambda_{1},\lambda_{2}}=<\Omega\lambda_{1}\lambda_{2}|M|JM>
\end{equation}
Helicity amplitude $A$ can be divided into canonical state expansions:
\begin{equation}
\begin{split}
A_{\lambda_{1},\lambda_{2}}
&=\sum_{m_{1}m_{2}}D^{s_{1}*}_{m_{1}\lambda_{1}}(\Omega)D^{s_{2}*}_{m_{2}-\lambda_{2}}(\Omega)A_{m_{1},m_{2}}
\end{split}
\end{equation}
We then proceed to decompose the partial wave of orbital angular momentum.
\begin{equation}
|lmsm_{s}>=\int d\Omega Y^{l}_{m}(\Omega)|\Omega sm_{s}>
\end{equation}
Therefore, the following results of helicity state can be obtained:
\begin{equation}
A_{m_{1},m_{2}}=\sum_{lsmm_{s}}(lmsm_{s}|JM)Y^{l}_{m}(\Omega)G_{ls}^{j}
\end{equation}
Of which $G_ {ls}^{j} (lmsm_{s}|jm) = (lmsm_{s}|\boldsymbol{M}|jm) $, and then according to the property of $D$function and  $CG$ coefficient.
Finally we can get the following partial wave expansion.
To get partial amplitude ,we have usen the orthogonality of $CG$ coefficients
\begin{equation}
\begin{split}
&\sum_{m_{1}m_{2}}(j_{1}m_{1}j_{2}m_{2}|jm)(j_{1}m_{1}j_{2}m_{2}|j^{\prime}m^{\prime})=\delta_{jj^{\prime}}\delta_{mm^{\prime}}\\
&\sum_{jm}(j_{1}m_{1}j_{2}m_{2}|jm)(j_{1}m_{1}^{\prime}j_{2}m_{2}^{\prime}|jm)=\delta_{m_{1}m_{1}^{\prime}}\delta_{m_{2}m_{2}^{\prime}}
\end{split}
\end{equation}
Then the partial wave scattering amplitude can be represented as follows,where $N_{J}=\sqrt{\frac{2J+1}{4\pi}}$.\\
\begin{align}
&A_{\lambda_{1},\lambda_{2}}
=N_{J}F^{J}_{\lambda_{1}\lambda_{2}}D^{J*}_{M\lambda}(\phi,\theta,0)
\end{align}
The helicity amplitude can be expanded by partial wave amplitudes.
\begin{equation}
\begin{split}
F^{J}_{\lambda_{1}\lambda_{2}}=\sum_{ls}\sqrt{\frac{2l+1}{2J+1}}a^{J}_{ls}(l0s\lambda|J\lambda)(s_{1}\lambda_{2}s_{2}-\lambda_{2}|s\lambda)
\end{split}
\end{equation}
Recall that in the helicity frame of a decaying particle,the helicity $\lambda$ of one particle is equal to $\pm m_{z}$ of that particle.So the state $|\Omega m_{1}m_{2}>$ equal to $|\Omega \lambda_{1},-\lambda_{2}>$ in helicity frame.And the helicity state of two particles can be represented as follows for particles with definite total angular momentum.And for a spin $\frac{1}{2}$ state,the normalization constant $N_{J}=\sqrt{\frac{2J+1}{4\pi}}$.Inversely the partial wave states $|JMls>$ can also be expressed by the two particle helicity states.

\section{A new  decomposition method}


For spin one particle,we can also do the following expansion easily.This is a key step for partial wave expansion.We can see that the helicity states of spin one with arbitrary momentum can be decomposed into canonical states in rest frame directly.This is why the helicity amplitude is proportional to a $D(R)$ function.We can write the helicity states of spin one and momentum $p=(p_{0},psin\theta cos\phi,psin\theta sin\phi,pcos\theta)$ as follows.
\begin{align}
&\boldsymbol{\epsilon}^{H}_{\pm}=\frac{1}{\sqrt{2(p_{x}^{2}+p_{y}^{2})}}(0,-\frac{p_{x}p_{y}}{|\boldsymbol{p}|}\pm ip_{y},-\frac{p_{y}p_{z}}{|\boldsymbol{p}|}\mp ip_{x},\frac{p_{x}^{2}+p_{y}^{2}}{|\boldsymbol{p}|})\notag\\
&\;\;\;\;\;\;\;\;\;\;\;\;\;\;\;\;\;\;\;\;\;\;\;\;\boldsymbol{\epsilon}^{H}_{0}=(\frac{|\boldsymbol{p}|}{m},\frac{\boldsymbol{p}}{|\boldsymbol{p}|}\frac{p_{0}}{m})
\end{align}

\begin{figure*}[!htb]
\centering
\includegraphics[width=0.45\textwidth,height=0.35\textwidth]{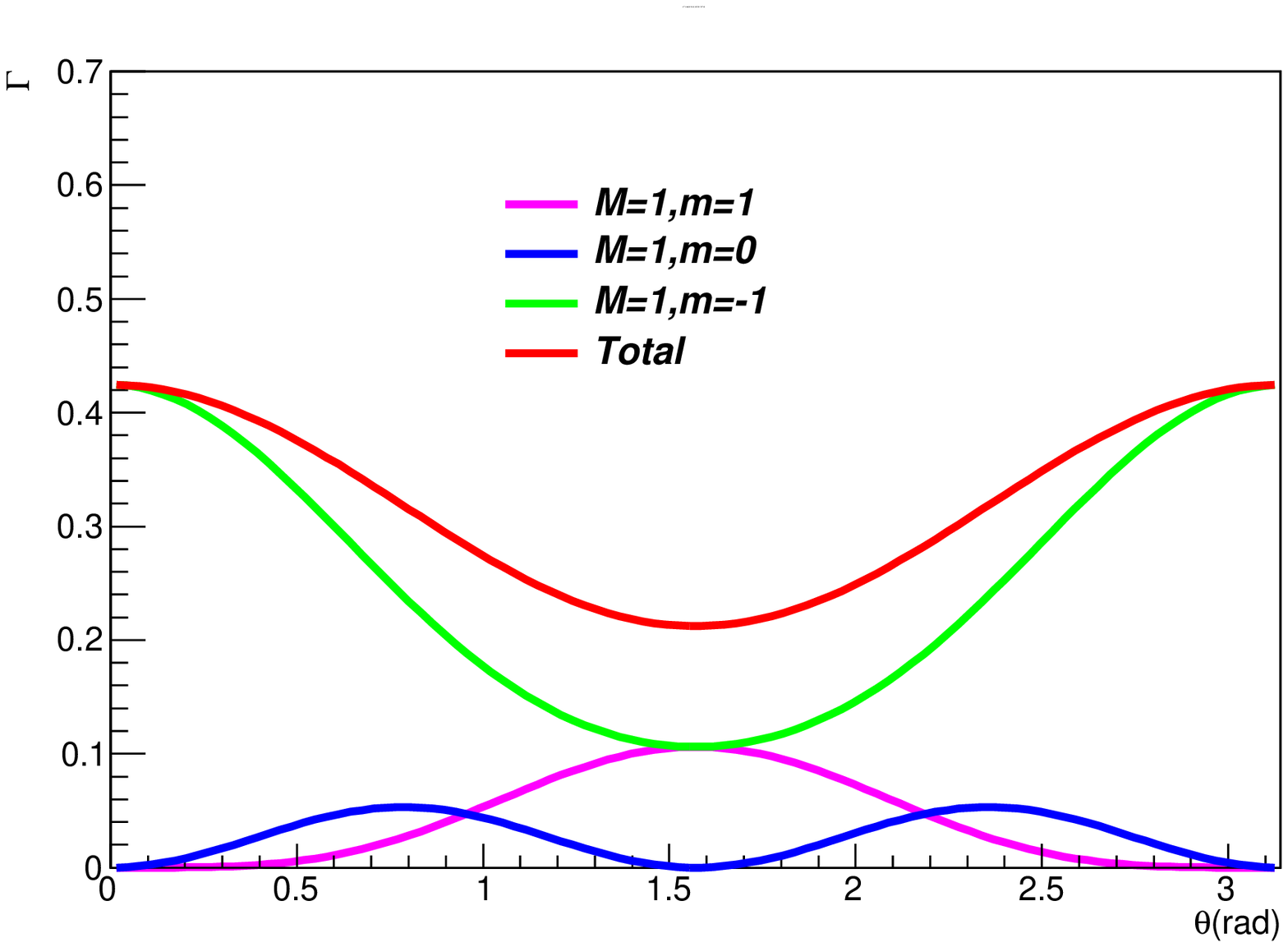}
\put(-21, 8){\textbf{ }}
\includegraphics[width=0.45\textwidth,height=0.35\textwidth]{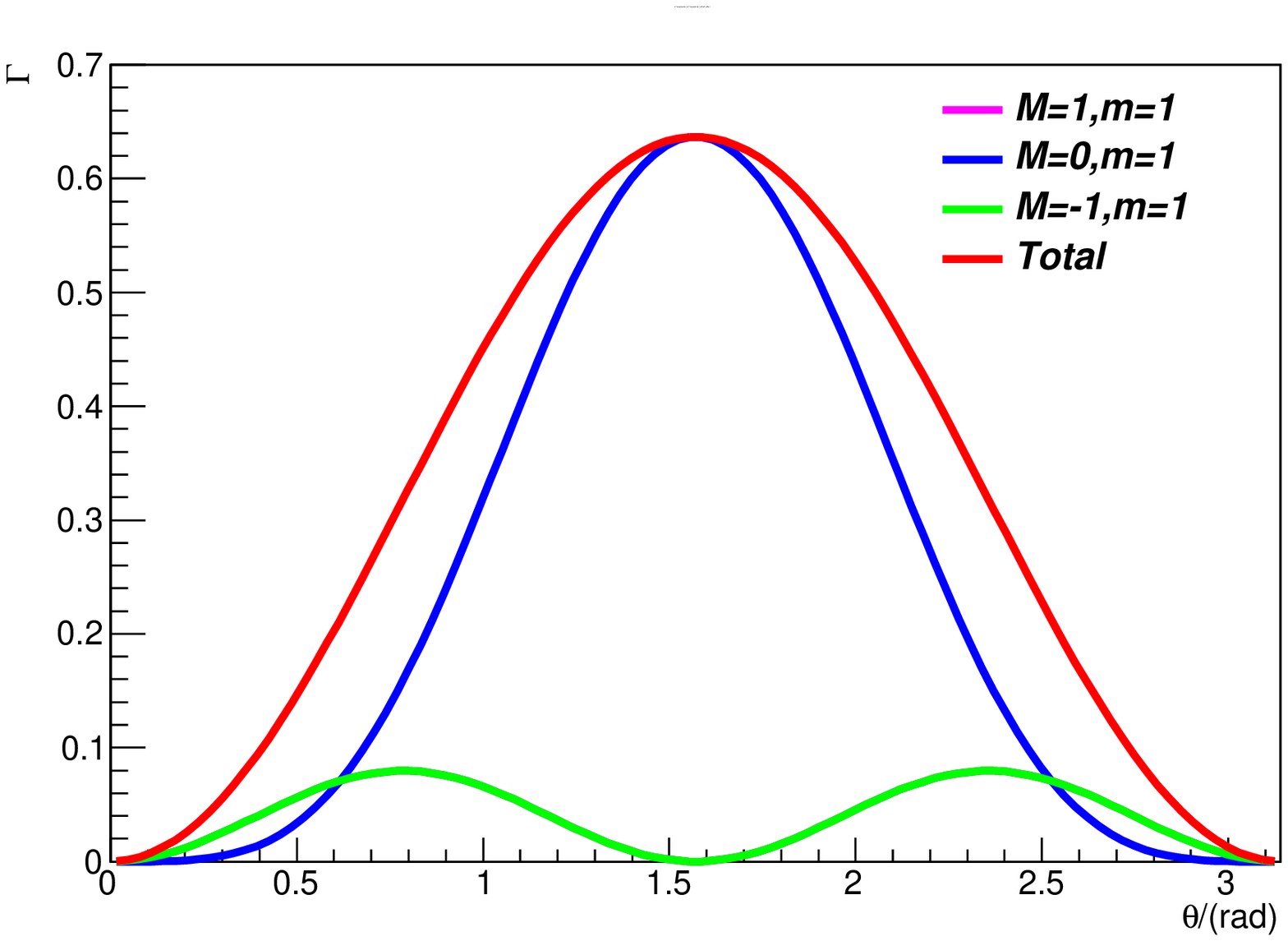}
\put(-21, 8){\textbf{ }}\\
\includegraphics[width=0.45\textwidth,height=0.35\textwidth]{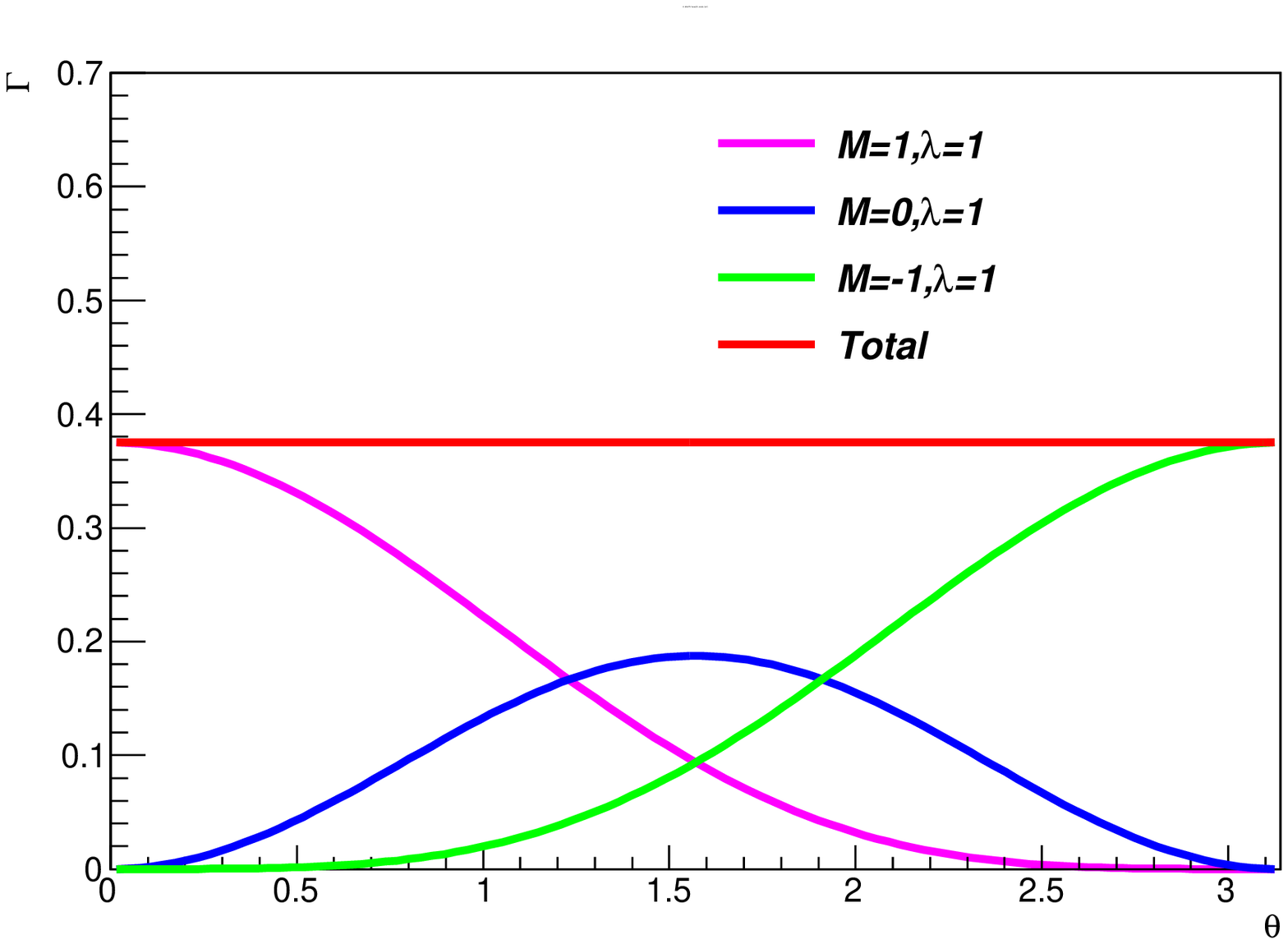}
\put(-21, 8){\textbf{ }}
\includegraphics[width=0.45\textwidth,height=0.35\textwidth]{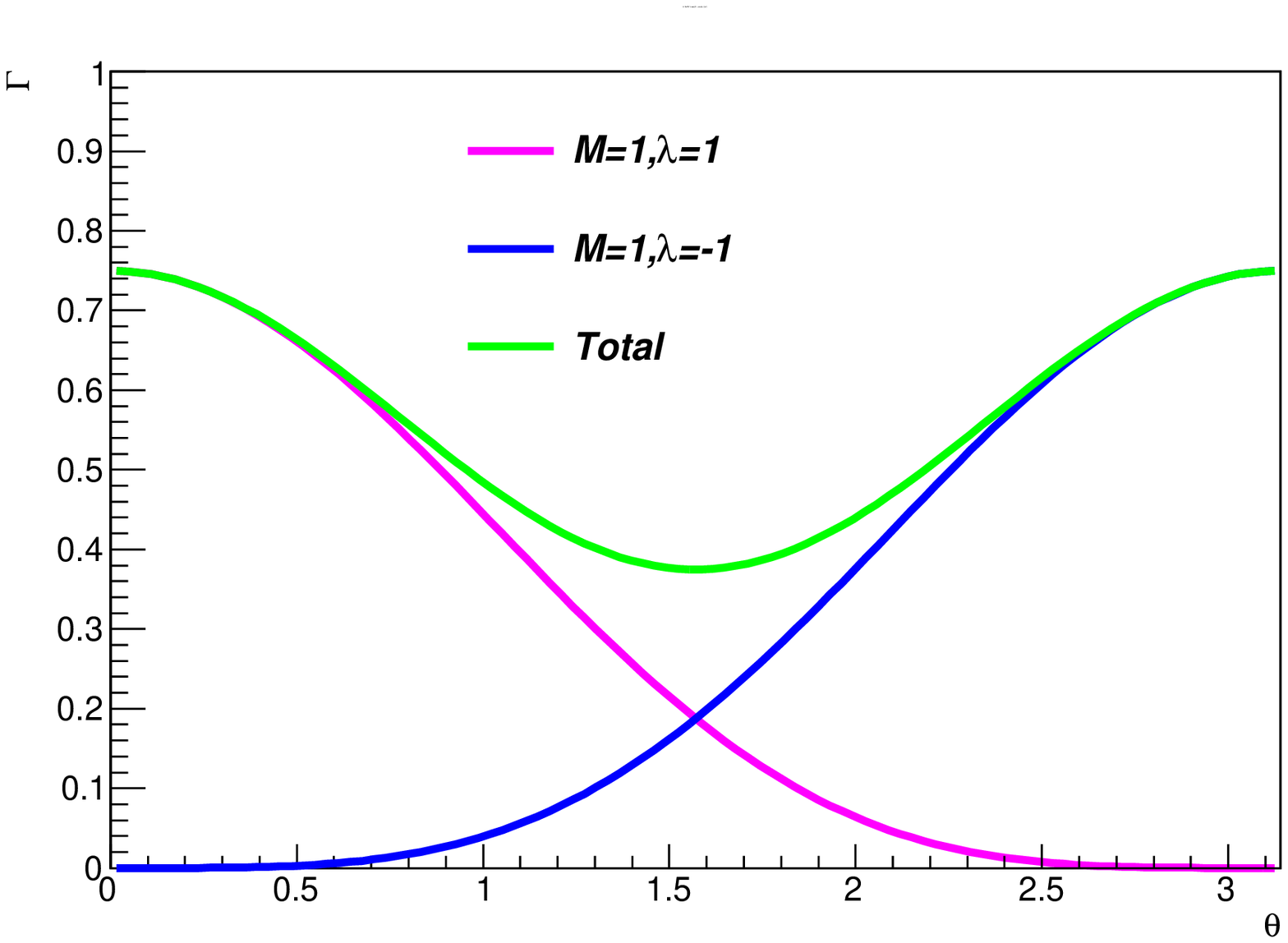}
\put(-21, 8){\textbf{ }}\\
\caption{ $V \rightarrow \bar{\psi} \psi$  decays to canonical states,the angular distribution vs $\theta$.}
\label{fig:ngam_thrust1}
\end{figure*}

This is the result of the rotational operator acting on the canonical states.And we can see that in this decomposition the canonical states do not depend on the angular part,so we can get the states do not depend on angular and a $D(R)$ function which depends on polar angle $\theta$,$\phi$.
\begin{equation*}
\begin{cases}
&\epsilon_{+}(\boldsymbol{p})=D^{1*}_{11}(R)\epsilon_{+}+D^{1*}_{01}(R)\epsilon_{0}+D^{1*}_{-11}(R)\epsilon_{-}\\
&\epsilon_{0}(\boldsymbol{p})=\frac{p_{0}}{m}D^{1*}_{10}(R)\epsilon_{+}+\frac{p_{0}}{m}D^{1*}_{00}(R)\epsilon_{0}+\frac{p_{0}}{m}D^{1*}_{-10}(R)\epsilon_{-}\\
&\epsilon_{-}(\boldsymbol{p})=-D^{1*}_{1-1}(R)\epsilon_{+}-D^{1*}_{0-1}(R)\epsilon_{0}-D^{1*}_{-1-1}(R)\epsilon_{-}\\
\end{cases}
\end{equation*}
And for states whose momentum is in the negative direction $-\boldsymbol{p}$ .
\begin{equation*}
\begin{cases}
&\epsilon_{+}(-\boldsymbol{p})=-D^{1*}_{1-1}(R)\epsilon_{+}-D^{1*}_{0-1}(R)\epsilon_{0}-D^{1*}_{-1-1}(R)\epsilon_{-}\\
&\epsilon_{0}(-\boldsymbol{p})=-\frac{p_{0}}{m}D^{1*}_{10}(R)\epsilon_{+}-\frac{p_{0}}{m}D^{1*}_{00}(R)\epsilon_{0}-\frac{p_{0}}{m}D^{1*}_{-10}(R)\epsilon_{-}\\
&\epsilon_{-}(-\boldsymbol{p})=D^{1*}_{11}(R)\epsilon_{+}+D^{1*}_{01}(R)\epsilon_{0}+D^{1*}_{-11}(R)\epsilon_{-}\\
\end{cases}
\end{equation*}
So we can get the result of rotational operator acting on the canonical states.
\begin{equation}
U(R)|JM\lambda>=\sum_{M^{\prime}}D^{j*}_{M^{\prime}M}(R)|JM^{\prime}\lambda>
\end{equation}
Next we prove the amplitude do not depend on $m$,or it possesses rotational symmetry.First if this process has rotational symmetry,then the relation holds $U(R)\boldsymbol{T}U(R)^{-1}=\boldsymbol{T}$.Then the rotational operator $U(R)$ acts on initial state and final state.
\begin{equation}
\begin{split}
&<JM|\boldsymbol{T}U(R)|JM^{\prime}\lambda>=D^{j*}_{M^{\prime}M}(R)<JM^{\prime}|\boldsymbol{T}|JM^{\prime}\lambda>\\
&<JM|U(R)\boldsymbol{T}|JM^{\prime}\lambda>=D^{j*}_{M^{\prime}M}(R)<JM|\boldsymbol{T}|JM\lambda>
\end{split}
\end{equation}
Finally,compare the above two relations,we can get the following equation \ref{JM}.
\begin{equation}
\label{JM}
<JM^{\prime}|\boldsymbol{T}|JM^{\prime}\lambda>=<JM|\boldsymbol{T}|JM\lambda>
\end{equation}
Then we consider a special example:a vector particle decaying into two massless fermions $V\rightarrow \bar{\psi}\psi$.The interaction has the form $V^{\mu}\bar{\psi}\gamma_{\mu}\psi$.We only consider the left handed $l^{-}$ and right handed $l^{+}$ for helicity conservation.For two particle states,we can get the two-particle helicity states as follows,which can be expressed by canonical states.As  \autoref{fig:ngam_thrust1} shown is the canonical and helicity amplitudes of $V\rightarrow \bar{\psi}\psi$.

As studied previously,helicity amplidude can be decomposed into partial waves.
For one spin zero particle decays into two fermions,only $S$ wave and $D$ wave can contribute.
Next we will study the partial wave amplitude,and get the relation between helicity amplitude and partial wave amplitude.
According to previous partial wave analysis,We can get the following relation.
\begin{equation}
F^{J}_{\lambda}=\sum_{ls}(\frac{2l+1}{2J+1})^{\frac{1}{2}}a_{ls}^{J}(l0s\lambda|J\lambda)
\end{equation}
Obviously, the partial wave amplitude $G_{LS} $ and helicity amplitude $H_{\lambda}$ can be related through a transformation of  unitary matrix $C$, which can be expressed as follows.
\begin{equation}
\begin{split}
&\overrightarrow{H}_{\lambda}=C\cdot  \overrightarrow{G}_{L}\;\;\;\;\;\overrightarrow{G}_{L}=C^{-1}\cdot  \overrightarrow{H}_{\lambda}
\end{split}
\end{equation}
Where the vector $\overrightarrow{H}_{\lambda}=(H_{\lambda},...,H_{\lambda^{\prime}})$,and $\overrightarrow{G}_{L}=(G_{L},...,G_{L^{\prime}})$,The value of $L$ is $S$,$P$and $D$.For a decay process $\frac{1}{2}\rightarrow \frac{1}{2}+0$,The inverse of the transformation matrix $C$ and itself is as follows.
\begin{equation}
\begin{split}
&C=\begin{pmatrix}
\sqrt{\frac{1}{2}} & -\sqrt{\frac{1}{2}}  \\
\sqrt{\frac{1}{2}} & \sqrt{\frac{1}{2}}
\end{pmatrix}
\;\;\;\;C^{-1}=\begin{pmatrix}
\sqrt{\frac{1}{2}} & \sqrt{\frac{1}{2}}  \\
-\sqrt{\frac{1}{2}} & \sqrt{\frac{1}{2}}
\end{pmatrix}
\end{split}
\end{equation}
Then the amplitude with orbital angular momentum of $0$and $1$ can be obtained from the matrix $C$,$H_{+}$and $H_{-}$.
\begin{equation}
\begin{split}
&G_{S}=\sqrt{\frac{1}{2}}H_{+}+\sqrt{\frac{1}{2}}H_{-}\\
&G_{P}=-\sqrt{\frac{1}{2}}H_{+}+\sqrt{\frac{1}{2}}H_{-}
\end{split}
\end{equation}
\indent 
Another practical example is $1\rightarrow 1+0$. Generally, according to the theory of angular momentum coupling, there may be contributions of $S$, $P$and $D$waves.Assuming parity is conserved in the  process, when the product of intrinsic parity of all particles is negative, so there is only $P$wave amplitude. However when the product of intrinsic parity of all particles is positive, there can only be $D$and $S$wave amplitude contributions. The transformation matrix of partial wave amplitude and helicity state amplitude is as follows.
\begin{equation}
\begin{split}
&C=\begin{pmatrix}
\sqrt{\frac{1}{3}} & -\sqrt{\frac{1}{2}} & \sqrt{\frac{1}{6}} \\
\sqrt{\frac{1}{3}} & 0 & -\sqrt{\frac{2}{3}} \\
\sqrt{\frac{1}{3}} & \sqrt{\frac{1}{2}} & \sqrt{\frac{1}{6}} 
\end{pmatrix}
\end{split}
\end{equation}
And recall that in the helicity frame of a decaying particle,the helicity\cite{2000general} $\lambda$ of one particle is equal to $\pm m_{z}$ of that particle.
In order to study the partial wave in a decay process,we have to lead in projection operator.Because projection operator can help us to get the pure orbital momentum states.Ordinarily a wave can have a mixing orbital contribution.So it.s necessary to introduce the projection operators.
The spin operator after  boost in a specific direction is called "moving spin" in some literatures.So, how to construct the wave function with higher spin? In fact, it can be obtained by the method of angular momentum coupling. 
Therefore, for the two body decay process $1\rightarrow 1+0$, we can conclude that the relationship between the partial wave amplitude and the helicity amplitude is as follows (under the low energy limit $\frac{q_{0}}{m}\rightarrow 1$).
\indent 
it can be seen from the above equation, if parity of the decay process is conserved, $H_ {\lambda}=\pm H_{-\lambda}$. So when $H_ {\lambda}=- H_{-\lambda}$, we can conclude that there is only $P$wave in the decay process, and the contribution of $S$wave and $D$wave is zero. And if $H_ {\lambda}=H_{-\lambda}$, there are only $S$ wave and $D$ wave,  the contribution of $P$ wave is zero. 
For example, a spin wave function with a spin of $J$ can be get by coupling $J$ wave functions with spin $1$.
\begin{equation*}
\begin{split}
&\phi^{j}_{M}(\boldsymbol{p})=\sum_{m_{i}}\Pi_{k=1}^{k=j}(km_{k}^{\prime}1m_{k+1}|k+1 m_{k+1}^{\prime})\phi^{1}_{m_{k}}(\boldsymbol{p})
\end{split}
\end{equation*}
Similarly, the helicity wave function with spin of $J$ can also be obtained by coupling $J$  wave functions with  spin of $1$,only by replacing $m_ {i}$ by $ \lambda_ {i}$.
\begin{equation*}
\begin{split}
&\phi^{j}_{\Lambda}(\boldsymbol{p})=\sum_{\lambda_{i}}\Pi_{k=1}^{k=j}(k\lambda_{k}^{\prime}1\lambda_{k+1}|k+1 \lambda_{k+1}^{\prime})\phi^{1}_{\lambda_{k}}(\boldsymbol{p})\\
\end{split}
\end{equation*}

\section{Projection operator}
The spin summation can be written as the following form.
\begin{equation}
\sum_{\lambda}\epsilon^{\mu}_{\lambda}\epsilon_{\lambda}^{*\nu}=(-g^{\mu\nu}+\frac{p^{\mu}p^{\nu}}{M^{2}})
\end{equation}
The projection operators for different angular momentum $J$ is as follows,where $W$ is the particle mass.
\begin{equation}
\begin{split}
&P_{\alpha;\beta}
=\widetilde{g}_{\alpha\beta}=-g_{\alpha\beta}+\frac{p_{\alpha}p_{\beta}}{W^{2}}\\
&P_{\alpha\beta;\gamma\delta}
=\frac{1}{2}(\widetilde{g}_{\alpha\gamma}\widetilde{g}_{\beta\delta}+\widetilde{g}_{\alpha\delta} \widetilde{g}_{\beta\gamma})-\frac{1}{3}\widetilde{g}_{\alpha\beta}\widetilde{g}_{\gamma\delta}
\end{split}
\end{equation}
Where the notation $(ab)$ and $(abcd)$ are defined  as follows.
\begin{equation}
\begin{split}
&\;\;\;\;\;(a\cdot b)=[a\cdot b]=g_{\alpha \beta}a^{\alpha}b^{\beta} \\    
&(abcd)=[abcd]=\epsilon_{\alpha\beta\gamma\delta}a^{\alpha}b^{\beta}c^{\gamma}d^{\delta}
\end{split}
\end{equation}

In fact, the relative angular momentum vector $\boldsymbol{r}$ can be expanded into the form of spherical harmonics, which are also first order irreducible tensor operators.
\begin{equation}
\begin{split}
&\boldsymbol{r}=\sqrt{\frac{4\pi}{3}}r\sum_{m}Y^{*}_{1m}(\theta,\varphi) \phi_{m}
\end{split}
\end{equation}
\indent 
Next, we verify whether the projection operator acts on the spin wave function and orbital wave function can lead to the wave function of a specific quantum number. We first calculate the result of the projection operator with orbital angular momentum of $1$ acts on the orbital wave function, that is, $\boldsymbol{r}$. Here, the $0$ component of the relative angular momentum is ignored, because it does not contribute any component under the action of the projection operator, so only the space-$3$ part need to be considered.Firstly, the projection operator with orbital angular momentum $1$ acts on the orbital wave function,we can get the following results:
\begin{equation}
\begin{split}
&P^{(1)}\boldsymbol{r}=\sum_{m}|1m><1m|\boldsymbol{r}=\sum_{m}Y^{*}_{1m}(\theta,\varphi)|jm>
\end{split}
\end{equation}
Similarly, when the projection operator with orbital angular momentum of $l=2$ acts on the orbital wave function, the orbital wave function with angular momentum $2$ can be obtained just multiplied by a constant factor.
\begin{align}
&\;\;\;\;P^{(2)}\boldsymbol{r}\boldsymbol{r}=\sum_{M}|2M><2M|\boldsymbol{r}\boldsymbol{r}\notag\\
&=\sum_{M}|2M><2M|Y^{*}_{1m_{1}}(\theta,\varphi)Y^{*}_{1m_{2}}(\theta,\varphi)|1m_{1}>|1m_{2}>\notag\\
&=\sum_{M}|2M><2M|\sum_{j^{\prime}M^{\prime}}|2M^{\prime}>Y^{*}_{j^{\prime}M^{\prime}}(\theta,\varphi)(j0|1010)\notag\\
&=\sum_{M}(20|1010)Y^{*}_{2M}(\theta,\varphi)|2M>
\end{align}
The following result is the result of the projection operator with orbital angular momentum  acting on the orbital wave function.
\begin{align}
&P^{(3)}\boldsymbol{r}\boldsymbol{r}\boldsymbol{r}=\sum_{M}|3M><3M|\boldsymbol{r}\boldsymbol{r}\boldsymbol{r}\notag\\
&=\sum_{M}(30|2010)(20|1010)Y^{*}_{3M}(\theta,\varphi)|3M>
\end{align}
\indent
When the projection operator acts on the orbital wave function, the wave function of specific angular momentum can be obtained. Similarly, in order to obtain the wave function of a specific spin, the spin projection operator is also required to act on the spin wave function. By using the relationship between the helicity state and the regular canonical state, the effect of the projection operator with spin $1$ on the spin wave function can be obtained.
\begin{equation}
\begin{split}
&\sum_{\lambda}\phi(\boldsymbol{p})_{\lambda}=\sum_{\lambda m}D^{j*}_{m\lambda}(\theta,\phi)|1m>
\end{split}
\end{equation}
The spin wave function with a spin of $J$ is as follows.
\begin{equation}
\begin{split}
&P^{(J)}=\sum_{M}|JM><JM|\\
\end{split}
\end{equation}
Similarly, the effect of the projection operator with spin $2$ on spin wave function is the following result.
From the above partial wave expansion, it can be seen that the helicity state with angular momentum of $j$in any momentum direction can be expanded into a series of canonical states without angular part multiplied by the corresponding $D(R)$ function. The difference is only that the constant $C_ {j}$ multiplied by the state depends on the helicity $\lambda$ of the particle.

\section{Polarization in density matrix }
The density matrix of the initial particle is as follows.The unpolarized part is just a unit matrix.
\begin{equation}
\begin{split}
&\rho=\rho_{0}+\sum_{L=1}^{2j}\rho^{(L)}\;\;\;\;\;\;\;\;\rho_{0}=\frac{1}{2j+1}\boldsymbol{1}
\end{split}
\end{equation}
The polarized parts are matrixes as follows.
\begin{equation}
\begin{split}
&\rho^{(L)}=\frac{2j}{2j+1}\sum_{M=-L}^{M=L}Q_{M}^{(L)}T_{M}^{(L)}\\
&(T_{M}^{(L)})^{m}_{n}=(2j+1)^{1/2}(jmLM|jn)
\end{split}
\end{equation}
\begin{table*}[!hbt]
\centering
\caption{The transformation between physical basis and $|jm><jm^{\prime}|$ basis}
\label{pj}
\begin{tabular}{cccccccccc}
\hline
Basis & $(++)$ &  $(00)$ &  $(--)$ &  $(+0)$& $(+-)$&  $(0+)$& $(0-)$& $(-+)$& $(-0)$\\
\hline
$|x+><x+|$ & $\frac{1}{4}$ &$\frac{1}{2}$&$\frac{1}{4}$&$-\frac{1}{2\sqrt{2}}$&$-\frac{1}{4}$&$-\frac{1}{2\sqrt{2}}$&$\frac{1}{2\sqrt{2}}$&$-\frac{1}{4}$&$\frac{1}{2\sqrt{2}}$\\
$|x\;\;0><x\;\;0|$ &$\frac{1}{2}$&$0$&$\frac{1}{2}$&$0$&$\frac{1}{2}$&$0$&$0$&$\frac{1}{2}$&$0$\\
$|x-><x-|$ &$\frac{1}{4}$ &$\frac{1}{2}$&$\frac{1}{4}$&$\frac{1}{2\sqrt{2}}$&$-\frac{1}{4}$&$\frac{1}{2\sqrt{2}}$&$-\frac{1}{2\sqrt{2}}$&$-\frac{1}{4}$&$-\frac{1}{2\sqrt{2}}$\\
$|y+><y+|$ &$\frac{1}{4}$ &$\frac{1}{2}$&$\frac{1}{4}$&$\frac{1}{2\sqrt{2}}$&$\frac{1}{4}$&$\frac{1}{2\sqrt{2}}$&$\frac{1}{2\sqrt{2}}$&$\frac{1}{4}$&$\frac{1}{2\sqrt{2}}$ \\
$|y\;\;0><y\;\;0|$ &$\frac{1}{2}$&$0$&$\frac{1}{2}$&$0$&$-\frac{1}{2}$&$0$&$0$&$-\frac{1}{2}$&$0$\\
$|y-><y-|$ &$\frac{1}{4}$ &$\frac{1}{2}$&$\frac{1}{4}$&$-\frac{1}{2\sqrt{2}}$&$\frac{1}{4}$&$-\frac{1}{2\sqrt{2}}$&$-\frac{1}{2\sqrt{2}}$&$\frac{1}{4}$&$-\frac{1}{2\sqrt{2}}$ \\
\hline
\end{tabular}
\end{table*}

Then we have to find the relation between the basis $|jm><jm^{\prime}|$ and irreducible tensors $T^{(L)}_{M}$.We list the transformation coefficients as shown in table \ref{wer}.
\begin{table*}[!hbt]
\centering
\caption{The transformation between physical basis and $|jm><jm^{\prime}|$ basis}
\label{wer}
\begin{tabular}{cccccccccc}
\hline
Basis & $(++)$ &  $(00)$ &  $(--)$ &  $(+0)$& $(+-)$&  $(0+)$& $(0-)$& $(-+)$& $(-0)$\\
\hline
$T^{(0)}_{0}$ & $\sqrt{\frac{1}{3}}$ &$\sqrt{\frac{1}{3}}$&$\sqrt{\frac{1}{3}}$&$$&$$&$$&$$&$$&$$\\
$T^{(1)}_{1}$ &$$&$$&$$&$-\sqrt{\frac{1}{2}}$&$$&$-\sqrt{\frac{1}{2}}$&$$&$$&$$\\
$T^{(1)}_{0}$ &$\sqrt{\frac{1}{2}}$ &$$&$-\sqrt{\frac{1}{2}}$&$$&$$&$$&$$&$$&$$\\
$T^{(1)}_{-1}$ &$$ &$$&$$&$\sqrt{\frac{1}{2}}$&$$&$$&$\sqrt{\frac{1}{2}}$&$$&$$ \\
$T^{(2)}_{2}$ &$$&$$&$$&$$&$$&$$&$$&$1$&$$\\
$T^{(2)}_{1}$ &$$ &$$&$$&$$&$$&$-\sqrt{\frac{1}{2}}$&$$&$$&$\sqrt{\frac{1}{2}}$ \\
$T^{(2)}_{0}$ &$\sqrt{\frac{1}{6}}$&$-\sqrt{\frac{2}{3}}$&$\sqrt{\frac{1}{6}}$&$$&$$&$$&$$&$$&$$\\
$T^{(2)}_{-1}$&$$ &$$&$$&$\sqrt{\frac{1}{2}}$&$$&$$&$-\sqrt{\frac{1}{2}}$&$$&$$ \\
$T^{(2)}_{-2}$&$$ &$$&$$&$$&$1$&$$&$$&$$&$$ \\
\hline
\end{tabular}
\end{table*}

If we know the polarization of one particle in one fixed direction,how can we get the polarization in arbitrary direction.For density matrix of spin $\frac{1}{2}$.we can find the relation of helicity state and canonical state.Their $Z$ axis defination is different just by a rotation with direction momentum $\boldsymbol{p}$.
\begin{equation}
\begin{split}
&|+>=D_{++}^{\frac{1}{2}}(R)|+^{\prime}>+D_{-+}^{\frac{1}{2}}(R)|-^{\prime}> \\
&|->=D_{--}^{\frac{1}{2}}(R)|-^{\prime}>+D_{+-}^{\frac{1}{2}}(R)|+^{\prime}>
\end{split}
\end{equation}
Assume oringinal density matrix is $\rho$.Then we can get the density matrix after rotation is $\rho^{\prime}=D \rho D^{+}$.
\begin{equation*}
\begin{split}
&\rho^{\prime}=\begin{pmatrix}
D_{++}^{\frac{1}{2}}(R) &  D_{+-}^{\frac{1}{2}}(R) \\ D_{-+}^{\frac{1}{2}}(R) &  D_{--}^{\frac{1}{2}}(R)
\end{pmatrix}
\rho
\begin{pmatrix}
D_{++}^{\frac{1}{2}*}(R) &  D_{-+}^{\frac{1}{2}*}(R) \\  D_{+-}^{\frac{1}{2}*}(R) &  D_{--}^{\frac{1}{2}*}(R)
\end{pmatrix}
\end{split}
\end{equation*}
Or we can express the new density matrix $\rho^{\prime}$ as follows.
\begin{equation}
\begin{split}
&\rho^{\prime}_{\Lambda \Lambda^{\prime}}=\sum_{\lambda\lambda^{\prime}}\rho_{\lambda\lambda^{\prime}}D^{j}_{\Lambda \lambda}(R)D^{j*}_{\Lambda^{\prime} \lambda^{\prime}}(R)
\end{split}
\end{equation}
Then for density matrix of a spin $\frac{1}{2}$ particle,we can simplified the density matrix if parity is conserved.
\begin{equation}
\begin{split}
&\rho=\begin{pmatrix}
\rho_{\frac{1}{2},\frac{1}{2}} & \rho_{\frac{1}{2},-\frac{1}{2}}\\
\rho_{-\frac{1}{2},\frac{1}{2}}  & \rho_{-\frac{1}{2},-\frac{1}{2}}
\end{pmatrix}
=\begin{pmatrix}
a & b \\
-b & a
\end{pmatrix}
=\begin{pmatrix}
a & ic \\
-ic & a
\end{pmatrix}
\end{split}
\end{equation}
The polarization of a spin $\frac{1}{2}$ particle is defined as equation \ref{po}.
\begin{equation}
\begin{split}
\label{po}
&P_{0}=\rho_{11}+\rho_{22} \;\;\;\;\;\;\;\;\; P_{x}=\rho_{12}+\rho_{21}\\
&P_{y}=-i(\rho_{12}-\rho_{21}) \;\;\; P_{z}=\rho_{11}-\rho_{22}
\end{split}
\end{equation}
From above symmetry relation,we can see that 
\begin{equation}
\begin{split}
&P_{0}=2a\;\;\;\;P_{x}=0\;\;\;\;P_{y}=2c\;\;\;\;P_{z}=0
\end{split}
\end{equation}
For spin one ,we can also write the density matrix as follows if parity is conserved.
\begin{equation}
\begin{split}
&\rho=\begin{pmatrix}
\rho_{1,1} & \rho_{1,0} & \rho_{1,-1} \\
\rho_{0,1} & \rho_{0,0} & \rho_{0,-1} \\
\rho_{-1,1} & \rho_{-1,0} & \rho_{-1,-1} 
\end{pmatrix}
=\begin{pmatrix}
a &b & c \\
d & e & -d \\
c & -b & a
\end{pmatrix}
\end{split}
\end{equation}
\begin{equation}
\rho=\begin{pmatrix}
a &b & c \\
b^{*} & 1-2a & -b^{*} \\
c & -b & a
\end{pmatrix}
\end{equation}
So for parity conservation case,we can simplify the polarization for spin-one particle.it can be decomposed by a new basis matrix,which is Gelmann matrix (spin one),then we can calculate the polarization by taking the trace of density matrix.
\begin{equation}
\begin{split}
&P_{0}=1\;\;\;\;P_{1}=2Re(\rho_{1,2})\\
&P_{2}=-2Im(\rho_{1,2})\;\;\;\;P_{3}=3\rho_{11}-1\\
&P_{4}=2\rho_{1,3}\;\;\;\;P_{5}=0\\
&P_{6}=-2Re(\rho_{2,3})\;\;\;\;P_{7}=-2Im(\rho_{2,3})\\
&P_{8}=\frac{1}{\sqrt{3}}(1-3\rho_{1,1})
\end{split}
\end{equation}
Next if the density matrix is given,we need to find the physical polarization for $x$ ,$y$and $z$ polarization.Or find their relation with polarization in irreducible tensor basis.First we have to transform the physical basis $|1m_{x}><1m_{x}|$ and $|1m_{y}><1m_{y}|$ $(m_{x,y}=0,\pm1)$into the $Z$ basis $|1m_{z}><1m_{z}^{\prime}|$ for spin one.Where $|x+>$ represents the polarization direction is $x$ axis,and the $m_{x}=+1$.
\begin{equation}
\begin{split}
&|x+>=\frac{i}{\sqrt{2}}|10>-\frac{i}{2}|1+>+\frac{i}{2}|1->\\
&|x0>=\frac{1}{\sqrt{2}}|1+>+\frac{1}{\sqrt{2}}|1->\\
&|x->=\frac{-i}{\sqrt{2}}|10>-\frac{i}{2}|1+>+\frac{i}{2}|1->
\end{split}
\end{equation}
For the y direction spin polarization,we have the following relations.
\begin{equation}
\begin{split}
&|y+>=\frac{i}{\sqrt{2}}|10>+\frac{i}{2}|1+>+\frac{i}{2}|1->\\
&|y0>=\frac{-i}{\sqrt{2}}|1+>+\frac{i}{\sqrt{2}}|1->\\
&|y->=\frac{1}{\sqrt{2}}|10>-\frac{i}{2}|1+>-\frac{i}{2}|1->
\end{split}
\end{equation}
Next we have to transform above basis into nine basises $|jm><jm^{\prime}|+|jm^{\prime}><jm|$,and then into $T^{(L)}_{M}$.Or to calcualte the $9\times 9$ transformation matrix.The relation between physical basis and $|jm><jm^{\prime}|+|jm^{\prime}><jm|$ basis is shown in table\ref{pj}.

\section{Symmetry of two body decay}
The parity $P$ and time reversal $T$ operator are defined as follows.
\begin{equation}
\begin{split}
&\boldsymbol{P}:\boldsymbol{x}\rightarrow -\boldsymbol{x}\;\;\; \boldsymbol{p}\rightarrow -\boldsymbol{p}\;\;\;\boldsymbol{J}\rightarrow \boldsymbol{J}\\
&\boldsymbol{T}:\boldsymbol{x}\rightarrow \boldsymbol{x}\;\;\;\;\;\boldsymbol{p}\rightarrow -\boldsymbol{p}\;\;\;\;\boldsymbol{J}\rightarrow -\boldsymbol{J}
\end{split}
\end{equation}
\begin{figure*}[!htb]
\centering
\includegraphics[width=0.95\textwidth,height=0.45\textwidth]{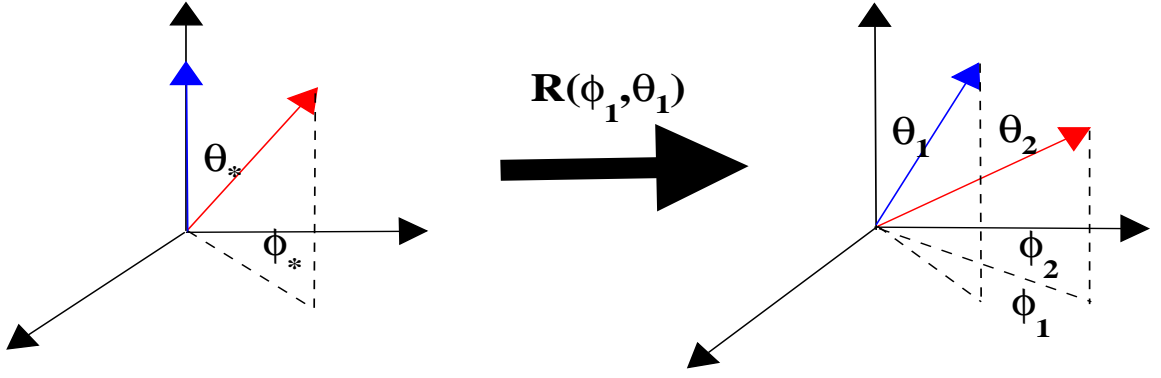}
\put(-21, 8){\textbf{ }}
\vspace{-10mm}
\caption{ The rotation of $R_{1}$}
\label{R1}
\end{figure*}
The canonical and helicity states transforms under P and T.
\begin{equation}
\begin{split}
&\Pi|\phi\theta p jm>=\eta|\pi+\phi, \pi-\theta jm>\\
&\Pi|\phi \theta, j \lambda>=\eta e^{-i\pi j}|\pi+\phi,\pi-\theta j -\lambda>\\
\end{split}
\end{equation}
We can get the parity transformation on orbital momentum states $\boldsymbol{r}$.
\begin{equation}
\begin{split}
&\boldsymbol{r}\rightarrow -\boldsymbol{r} \;\;\;\;\; |JMls>\rightarrow \eta (-1)^{l}|JMls>
\end{split}
\end{equation}
So we can get the transformation of orbital angular momentum $l$ state.
\begin{equation}
\begin{split}
&|lm>\rightarrow (-1)^{l}|lm>
\end{split}
\end{equation}
Next we will consider the symmetry of the density matrix under parity.It leads to the following symmetry relations.
\begin{equation}
\begin{split}
&\rho_{\lambda\lambda^{\prime}}^{J}=(-1)^{\lambda-\lambda^{\prime}}\rho_{-\lambda-\lambda^{\prime}}^{J}\;\;\rho_{mm^{\prime}}^{J}=(-1)^{m-m^{\prime}}\rho_{mm^{\prime}}
\end{split}
\end{equation}
We can get the symmetry relation.Assuming that the decay process occurs in the $X-Z$ plane, using the operator $\pi_ {y} = \pi e^ {-i\pi j_{y}}$ symmetry on helicity state:
\begin{equation}
\Pi_{y}|s\lambda>=\eta(-1)^{s-\lambda}|s-\lambda>
\end{equation}
Because of the nature of the $D$ function.
\vspace{0.15cm}
\begin{equation}
d^{j}_{m^{\prime}m}(\pi)=(-1)^{j-m}\delta_{m^{\prime}m}
\end{equation}
Then the density matrix under operation $\Pi_ {y}$.
\begin{align}
&\rho_{\lambda \lambda^{\prime}}=\sum_{MM^{\prime}}f_{M\lambda}D^{j*}_{M\lambda}(R)\rho_{MM^{\prime}}f_{M^{\prime}\lambda^{\prime}}^{*}D^{j}_{M^{\prime}\lambda^{\prime}}(R)\notag\\
&\;\;\rightarrow\sum_{MM^{\prime}}(-1)^{M-M^{\prime}}(-1)^{\lambda-\lambda^{\prime}}(-1)^{M-M^{\prime}}H_{-\lambda} \notag\\
&\;\;\;\;\; \times D^{j*}_{-M-\lambda}(R)\rho_{-M-M^{\prime}} H_{-\lambda^{\prime}}^{*}D^{j}_{-M^{\prime}-\lambda^{\prime}}(R)\notag\\
&\;\;=(-1)^{\lambda-\lambda^{\prime}}\rho_{-\lambda-\lambda^{\prime}}
\end{align}
Then, according to the properties of $D$ function, we can get that the properties of angular distribution under operation  $\Pi_ {y}$.
Next, we prove the second symmetry of the density matrix under the canonical state representation. Assuming that the reaction occurs in the $X-Y$ plane, we need to use the canonical state amplitude in $\Pi_ {z} = \pi e^ {-i\pi j_{z}}$transformation.
\vspace{0.15cm}
\begin{equation}
\Pi_{z}|sm>=\eta e^{-im\pi}|sm>
\end{equation}
Therefore, we can obtain the second symmetry of the density matrix, where we need to use the symmetry of the initial state density matrix:\;\;$\rho_{MM^{\prime}}=(-1)^{M-M^{\prime}}\rho_{MM^{\prime}}$.This symmetry is the result of the parity conservation.Next, we prove the symmetry of canonical states under parity transformation.
\begin{align}
\rho_{mm^{\prime}}&=\sum_{MM^{\prime}}f_{Mm}D^{j*}_{Mm}(R)\rho_{MM^{\prime}}f_{M^{\prime}m^{\prime}}^{*}D^{j}_{M^{\prime}m^{\prime}}(R)\notag\\
&\rightarrow\sum_{MM^{\prime}}(-1)^{M-m+m^{\prime}-M^{\prime}}D^{j*}_{Mm}(R)\rho_{MM^{\prime}}f_{M^{\prime}m^{\prime}}^{*}D^{j}_{M^{\prime}m^{\prime}}(R)\notag\\
&=\sum_{MM^{\prime}}(-1)^{m-m^{\prime}}D^{j*}_{Mm}(R)\rho_{MM^{\prime}}f_{M^{\prime}m^{\prime}}^{*}D^{j}_{M^{\prime}m^{\prime}}(R)\notag\\
&=(-1)^{m-m^{\prime}}\rho_{mm^{\prime}}
\end{align}

\section{Canonical state of Arbitrary polarization}
We can construct the canonical states of arbitrary polarization for spin one and spin $\frac{1}{2}$ particles.
\begin{equation}
|\boldsymbol{p}\boldsymbol{n}m>=L(\boldsymbol{p})U(\boldsymbol{n})|jm>
\end{equation}
Where the polar angle are defined as $\boldsymbol{p}=(\theta_{2},\phi_{2})$,and $\boldsymbol{n}=(\theta_{1},\phi_{1})$.
First we can define the rotation $R_{1}$ is a roatation which rotate the $Z$ axis to $\boldsymbol{n}$,and $R_{2}$ is a another rotation which bring $Z$ axis to momentum $\boldsymbol{p}$ direction.And we can define the relative rotation as follows.
\begin{equation}
R_{1}R_{*}=R_{2}\;\;\;U(R_{1})\boldsymbol{p}_{*}=\boldsymbol{p}
\end{equation}
We can do the following expansion.
\begin{equation}
\begin{split}
&|\boldsymbol{p}\boldsymbol{n}m>
=\sum_{\lambda}D^{*}_{m\lambda }(R_{*})|\boldsymbol{p}\lambda>\\
&=\sum_{\lambda k}D_{mk}^{*}(R_{1}^{-1})D_{k\lambda}^{*}(R_{2})|\boldsymbol{p}\lambda>
\end{split}
\end{equation}
In fact,we have the result.
\begin{equation}
\begin{split}
&|\boldsymbol{p}\boldsymbol{n}m>=\sum_{k}D_{km}(R_{1})|\boldsymbol{p}k>\\
&=\sum_{k\lambda}D_{km}(R_{1})D_{k\lambda}^{*}(R_{2})|\boldsymbol{p}\lambda>\\
&=\sum_{\lambda k}D_{mk}^{*}(R_{1}^{-1})D_{k\lambda}^{*}(R_{2})|\boldsymbol{p}\lambda>
\end{split}
\end{equation}
Where we have used the relation as follows.
\begin{equation}
D^{*}_{mk}(R_{1}^{-1})=D_{km}(R_{1})
\end{equation}
The next step is to calculate the canonical and helicity states with arbitrary polarization as follows.
\begin{equation}
\begin{split}
 &|\boldsymbol{p}\boldsymbol{n}m>=L(\boldsymbol{p})U(\boldsymbol{n})|m>\\
&\;\;\;\;=L(\boldsymbol{p})\sum_{k}D_{km}(R_{1})|k>\\
&\;\;\;\;=L(\boldsymbol{p})|\boldsymbol{n}m>
\end{split}
\end{equation}
As shown in \autoref{R1},that the direction of $\boldsymbol{p}_{*}$ is moved to $\boldsymbol{p}$ by the rotation $R_{1}$ action.

\section{Conclusion}
We have proposed a new method of decomposition of helicity waves,which can be shown that the decomposition contains two parts:one is just canonical states and the other  is $D(R)$ function.This decomposition is the reason why the decaying amplitude has the form of $H_{\lambda}D_{M\lambda}(R)$.Then we get the projection operator action on orbital and spin waves.Finally the symmetry of density matrix is also studied if parity is conserved,which has a close relationship with polarization.\cite{1999Muon,2010Basic,1992Tau,2019Polarization}
\bibliography{references}{}

\end{document}